\documentclass[aps,prb,preprint,superscriptaddress]{revtex4-1}

\usepackage{mathtext}
\usepackage{amssymb}
\usepackage{graphicx}
\usepackage{amsmath}

\begin{document}
\title{Energy relaxation rate of 2D hole gas in GaAs/InGaAs/GaAs quantum well within wide range of conductivitiy}

\author{I.~V.~Soldatov}
\author{A.~V.~Germanenko}
\affiliation{Ural State University, 620083 Ekaterinburg, Russia}
\author{G.~M.~Minkov}
\affiliation{Institute of Metal Physics RAS, 620990 Ekaterinburg,Russia}
\author{O.~E.~Rut}
\affiliation{Ural State University, 620083 Ekaterinburg, Russia}
\author{A.~A.~Sherstobitov}
\affiliation{Institute of Metal Physics RAS, 620990 Ekaterinburg,Russia}

\date{\today}

\begin{abstract}
The nonohmic conductivity of 2D hole gas (2DHG) in single $GaAsIn_{0.2}Ga_{0.8}AsGaAs$ quantum well structures within the temperature range of 1.4 - 4.2K, the carrier's densities  $p=(1.5-8)\cdot10^{15}m^{-2}$ and a wide range of conductivities $(10^{-4}-100)G_0$ ($G_0=e^2/\pi\,h$) was investigated. It was shown that at conductivity $\sigma>G_0$ the energy relaxation rate $P(T_h,T_L)$ is well described by the conventional theory (P.J. Price J. Appl. Phys. 53, 6863 (1982)), which takes into account scattering on acoustic phonons with both piezoelectric and deformational potential coupling to holes. At the conductivity range $0.01G_0<\sigma<G_0$ energy the relaxation rate significantly deviates down from the theoretical value. The analysis of $\frac{dP}{d\sigma}$ at different lattice temperature $T_L$ shows that this deviation does not result from crossover to the hopping conductivity, which occurs at $\sigma<10^{-2}$, but from the Pippard ineffectiveness.
\end{abstract}


\maketitle

\section{Introduction}
\label{sec:intr}

In case of conductivity over  delocalized states (diffusive conductivity) the electric field dependence of the conductivity originates from heating of 2DHG up to temperature $T_h$, greater than lattice temperature $T_L$. In the stationary conditions $T_h$ is determined by the balance between incoming energy rate $P _{in}$ and energy relaxation rate $P$. Therefore, studying the nonohmic conductivity, one can find the energy relaxation rate $P$, its dependence on $T_L$, $T_h$ and determine the main mechanisms of the energy relaxation. Moreover, the study of the nonohmic conductivity provides an opportunity to find the conditions in which the diffusion conductivity changes to hopping conductivity with the change of density, disorder and temperature. It is possible due to the fact that in the hopping regime of conductivity $\sigma(E)$ dependence results not only from the change of the carrier distribution over energy, but also from the change in probability of hopes along the field.

Over the past 20 years there have been published a number of experimental papers, investigating heating of 2D electron gas in GaAs structures at high conductivities \cite{Chow96,Chow97,Ma,Minkov01}, and a few papers studying p-type \cite{Gao,Gao2}, but only for lattice temperatures below 100mK. To the best of our knowledge, there is no experiment within a wide conductivity range at higher temperatures
In the present paper, we investigate the dependence of the energy relaxation rate on the carrier density and the strength of disorder in InGaAs based 2D hole structures in the temperature range of 1.4-4.2K and a wide range of conductivities $(10^{-4}-100)G_0$. We have obtained the following results. It was shown, that at conductivities above $G_0$ the energy relaxation rate is well described in terms of scattering on acoustic phonons \cite{Price}. At lower conductivities $(3\cdot10^{-2}-1)G_0$ the energy relaxation rate deviates down from the theory in Ref.~\onlinecite{Price}, while the regime of conductivity remains diffusive. This deviation is associated with the Pippard ineffectiveness of electron-phonon interactions. It was shown that a crossover to the hopping conductivity with lowering of $\sigma$ occurs at $\sigma\sim10^{-2}$.

\section{Experimental details}
The structures investigated were grown by metalorganic vapor-phase epitaxy on a semi-insulating GaAs substrate and consist of a 0.2mkm thick undoped GaAs buffer layer, a 10nm InGaAs quantum well and a 0.2-0.3mkm cup layer of undoped GaAs. The Carbone $\delta$-layer was situated at the distance of 7nm (samples 3855, 3857) or 15nm (3953) from each side of the quantum well. The samples were mesa-etched into the standard Hall bars. The hole density was varied by applying voltage to the Al gate electrode, deposited by thermal evaporation. Nonohmic conductivity measurements on gated structures require special care: the voltage drop along the sample must be significantly lower than the gate voltage, otherwise the distribution of the carrier under the gate electrode would be non-homogeneous. To avoid this effect the sample surface was covered by a 3-5mkm thick dielectric (parylen) layer before depositing the gate electrode. With this layer $\frac{dp}{dV_g}$ was less than $5\cdot10^{8}cm^{-2}V^{-1}$ and we could apply gate voltage $V_g$ up to 300V (while the voltage drop along the sample was less than 0.5V). The hole densities and conductivities at zero gate voltage and lattice temperature $T_L=1.4K$ for the structures investigated are listed in Table 1.

\begin{table}[ht]
\caption{The parameters of the samples.} 
\centering 
\begin{tabular}{c c c} 
\hline\hline 
Sample~~ & $~~p,\cdot10^{11}cm^{-2}~~$ & $~~\sigma(1.4K),G_0~~$ \\ [0.5ex] 
\hline 
3855 & 5.4 & 41 \\ 
3857 & 8.7 & 84 \\
3953 & 4 & 100 \\ [1ex] 
\hline 
\end{tabular}
\label{table:nonlin} 
\end{table}

\begin{figure}
\includegraphics[width=0.8\linewidth,clip=true]{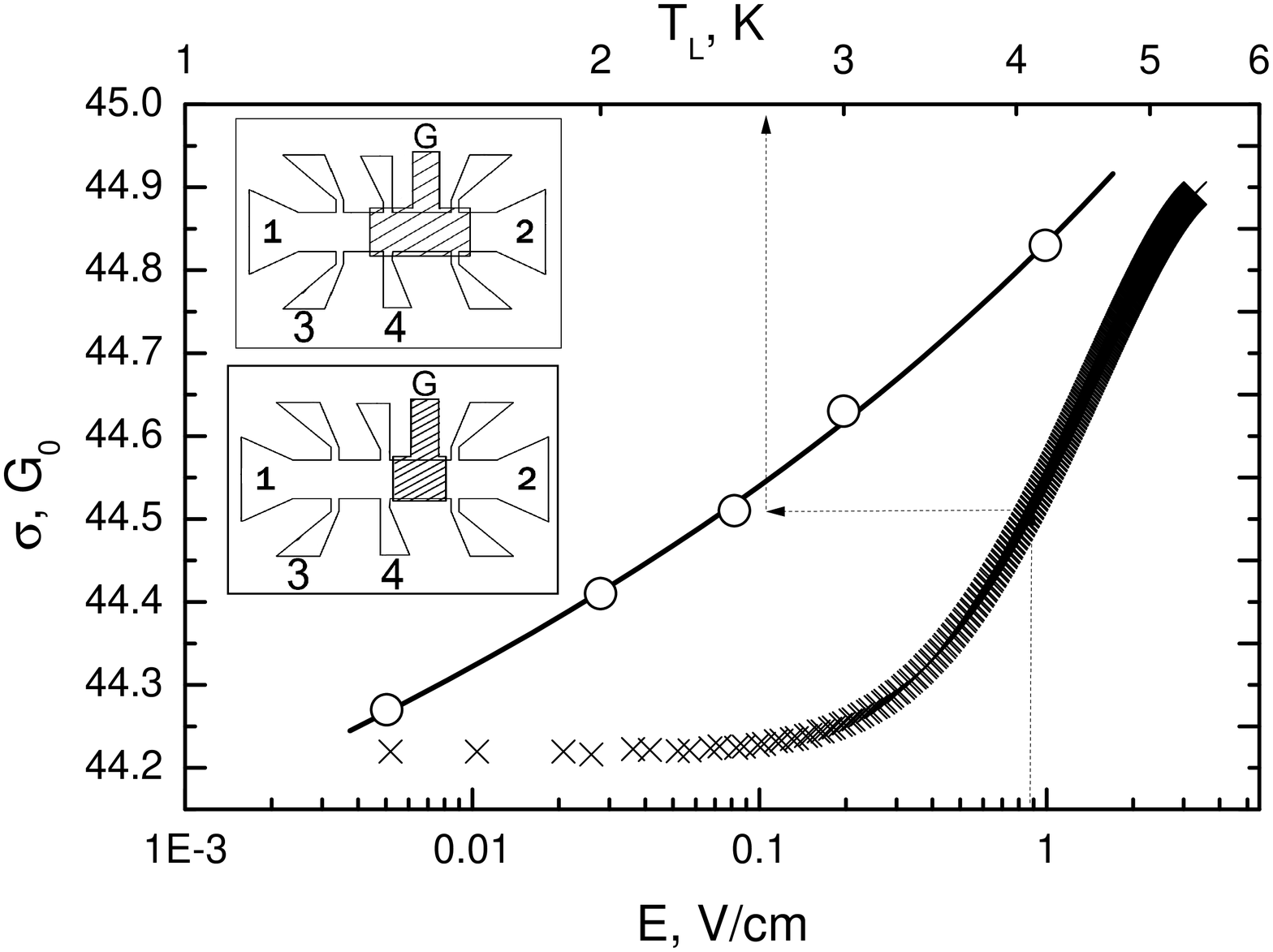}
\caption{The dependence of conductivity on lattice temperature in ohmic regime (open circles), and on electric field at $T_L=1.4K$ (crosses). The solid line is the approximation of $\sigma(E)$ by a smooth function. The upper inset - Hall bar with a gate for $p(V_g)$  measurements, lower inset - Hall bar with gate for heating experiments.
}\label{f1}
\end{figure}

The hole effective mass $m=0.16m_0$ was obtained from temperature dependence of Shubnikov-de Haas oscillations \cite{MinkovMass}. The dependence $p(V_g)$ was obtained in a set of measurements with a long gate electrode, covering two pairs of Hall contacts (upper inset in Fig.~\ref{f1}). The heating experiments were taken with a shorter (Hall contacts remained uncovered) gate electrode (lower inset in Fig.~\ref{f1}) to avoid the rise in contact resistance at a high bias voltage. The current dependence of voltage drop between potential contacts 3-4 was measured during the current sweep while the lattice temperature remained constant. Temperature dependence of conductivity was measured in a linear regime of response.

\section{Results and discussion}
\label{sec:results}

\begin{figure}
\includegraphics[width=0.8\linewidth,clip=true]{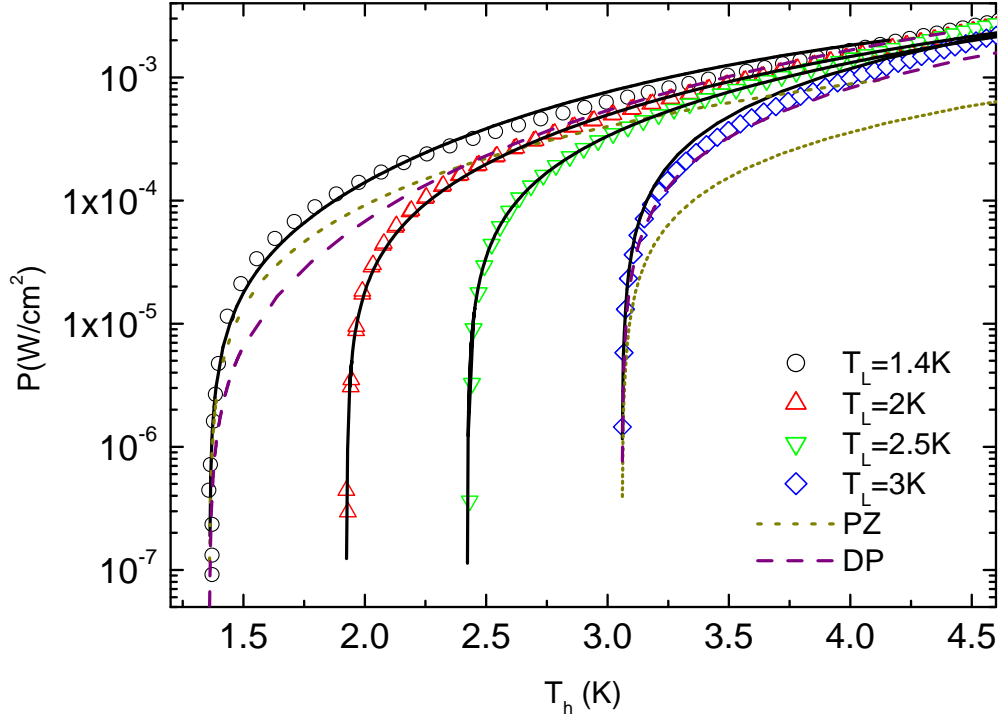}
\caption{(Color online) Hole temperature dependence of energy relaxation rate at different lattice temperatures for sample 3855 ($V_g=0, p=5.4\cdot 10^{11}, \sigma (1.4)=41.4G_0$). Dashed and dotted lines are the contributions from the coupling with deformation and piezoelectric potential, respectively, solid lines are their sums.
}\label{f2}
\end{figure}

Fig.~\ref{f1} shows the dependences of the conductivity on lattice temperature $\sigma(T_h)$ and electric field $\sigma(E)$ at lattice temperature $T_L=1.4K$ and $V_g=0$ for sample 3855. The temperature dependence of conductivity of degenerate gas ( $E_F\gg k_BT$, $E_F$- Fermi level) at low temperatures is fully determined by temperature dependence of quantum corrections to the conductivity. The corrections depend on nothing but the carrier temperature $T_h$ (the dependence $\sigma(T)$ for similar structures was investigated in Ref.~\onlinecite{Minkovsig}). In this case, having compared the temperature dependence of conductivity in ohmic regime and electric field dependence of conductivity, we have reconstructed (follow the dashed line in Fig. \ref{f1}) the electric field dependence of hole temperature $T_h$, and, then, we have calculated the incoming power $P_{in}=jE=\sigma E^2$, required to heat the holes up to $T_h$\cite{fnt1} . In stationary conditions $P_{in}$ is equal to energy relaxation rate $P(T_h,T_L)$.
The temperature dependence of the energy relaxation rate $P(T_h,T_L)$, obtained at different lattice temperature $T_L$ for sample 3855 is presented in Fig.~\ref{f2}. at conductivity 41.4$G_0$. Similar results were obtained for all the structures investigated over a wide conductivity range.

Let us compare our experimental data with the theory. As the mechanism governing a relaxation rate at low temperatures is scattering on acoustic phonons with both piezoelectric and deformational coupling, for the quantitative analysis we use the theory from Ref.~\onlinecite{Price}. The energy relaxation rate as a function of hole temperature $T_h$ and lattice temperature $T_L$ is written as a difference of two identical functions, where one term depends on $T_h$, and the other on $T_L$ (eq.(12) in Ref.~\onlinecite{Price}):

\begin{equation}
P(T_h,T_L)=\frac{n}{\hbar E_F}(F(T_h)-F(T_L))
 \label{eq1}
\end{equation}

with

\begin{equation}
F(T)=\int dq|I(q)|^2\overline{a(S(q))^2\cdot(\hbar w)^3\cdot N(\hbar w/k_BT)}
\end{equation}

where $I(q)=\int_{-d}^{d}\rho(z)e^{iqz}dz$ is the form factor for the normal-direction wave function $(\rho(z)=\sqrt{\frac{1}{d}}\cos(\frac{\pi z}{2d}))$; 2d is the width of quantum well; $S(q)$ is a screening factor; $N(\hbar w/k_BT)$ isthe phonon distribution function; $a$   is a quantity, associated with the three-dimensional scattering matrix.
In general $F(T)$ is the sum of deformation-coupled and piezoelectric-coupled contributions:

$F=F_{Ln}^{DP}+F_{Ln}^{PZ}+F_{Tr}^{PZ}$

where index Ln refers to the longitudinal component of the wave-vector, and Tr to the transversal one.
Then each component of $a$ could be expressed as follows\cite{Ma}:

\begin{equation}
a_{Tr}^{PZ}=b_0\frac{\alpha^2}{q^2+Q^2}\cdot \frac{8q^4Q^2+Q^6}{2(q^2+Q^2)^3}
\end{equation}

\begin{equation}
a_{Ln}^{PZ}+a_{Ln}^{DP}=b_0 \left(\frac{\alpha^2}{q^2+Q^2}\cdot \frac{9q^4Q^2+Q^6}{2(q^2+Q^2)^3}+1 \right)
\end{equation}

where $\alpha =\frac{eh_{14}}{\Xi}$, $h_{14}$ is a piezoelectric coupling constant, $\Xi$ - deformation potential, $q$ and $Q$ are the components of the wave-vector, respectively normal to and parallel to the heterolayer plane. The constant $b_0$ is equal to $\frac{m^*\Xi^2}{\hbar^22k_l}$ , where $k_l$ - is an elastic constant equal to  $\rho\cdot s_{Ln}^2$ ($\rho$ is the density of GaAs, $s_{Ln}^2$ is the longitudinal velocity of sound).

We have calculated temperature dependences of the energy relaxation rate using the following coupling constants: $h_{14}=1.5\cdot 10^9 V/m$ (Ref.~\onlinecite{Zook}), $\Xi =8eV$ (Ref.~\onlinecite{Notle}) and the two-dimensional screening constant $p=\left( \frac{a_B}{2}\right)^{-1}=0.5nm^{-1}$, where $a_B$ is an effective Bohr radius.

The calculated curves $P(T_h,T_L)$ are presented in Fig.~\ref{f2}. The dashed and dotted lines are the contributions of the deformation and piezoelectric coupling respectively. One can see that their contributions are comparable within the whole temperature range. The solid line is their sum. As seen from Fig.~\ref{f2} the theoretical curves are in good agreement with the experimental data.

It is impossible to compare directly our experimentally obtained values of the energy relaxation rate with the results in Refs. \onlinecite{Gao},\onlinecite{Gao2}, as they have been obtained at lower temperatures (below 100mK). However, if we calculate $P(T_h,T_L)$ with the coupling constants, we used, we would get the energy relaxation rate only twice bigger than the experimental value in Refs. \onlinecite{Gao},\onlinecite{Gao2}. We believe that it is quite a good agreement for extrapolation from units of Kelvin down to milli-Kelvin temperatures\cite{ftn2}.

\begin{figure}
\includegraphics[width=0.8\linewidth,clip=true]{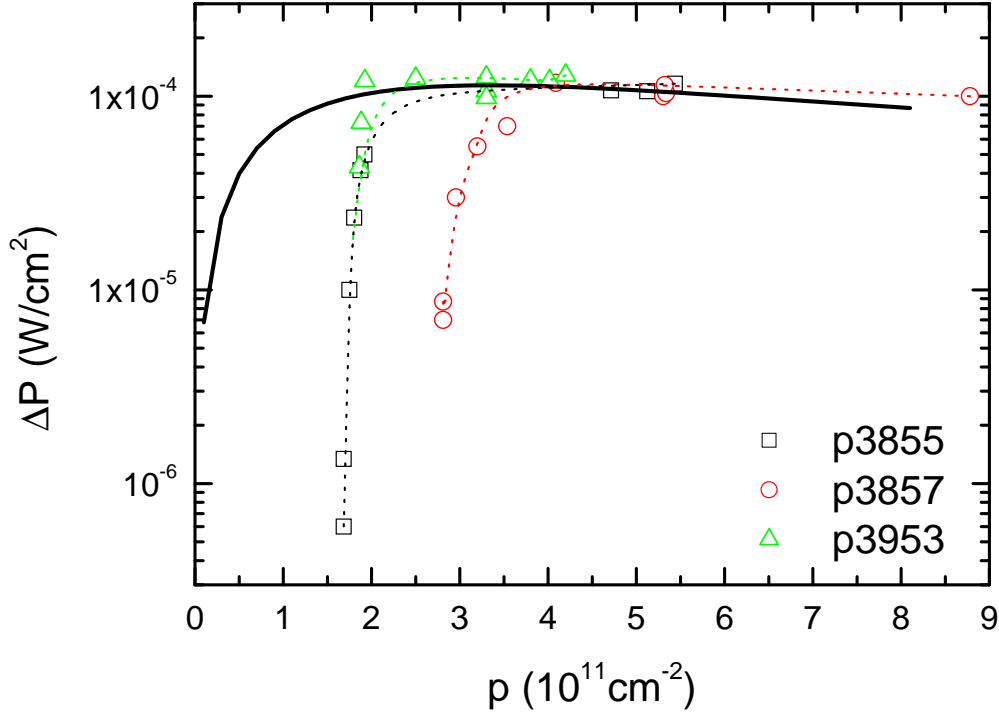}
\caption{(Color online) The power $\Delta P$ required to heat holes from lattice temperature $T_L=1.4K$ up to $T_h=1.9K$ as a function of concentration. The solid line is theoretical curve in accordance with Eq.~(\ref{eq1}). Doted lines are only a guide for an eye.
}\label{f3}
\end{figure}

\begin{figure}
\includegraphics[width=0.8\linewidth,clip=true]{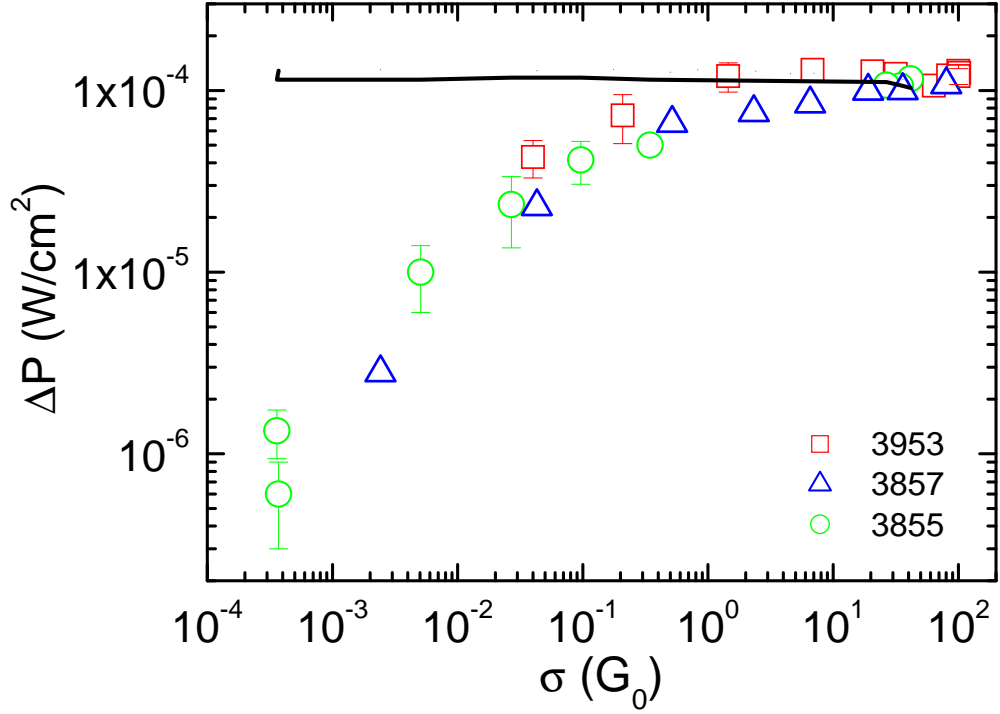}
\caption{(Color online) The power $\Delta P$ required to heat holes from lattice temperature $T_L=1.4K$ up to $T_h=1.9K$ as a function of conductivity at $T_L=1.4K$. Solid line is theoretical dependence $\Delta P(p(\sigma))$ for sample 3855.
}\label{f4}
\end{figure}

From Eq.~(\ref{eq1}) it follows that besides coupling constants, temperatures $T_h$ and $T_L$, the only sample parameter which the energy relaxation rate depends on is hole density. Let us analyze the dependence of the power $\Delta P$ required to heat holes from lattice temperature $T_L=1.4K$ up to $T_h=1.9K$, shown in Fig.~\ref{f3}. It is seen that at densities above a certain value (different for each sample), the experimental data is in agreement with the theory. At lower densities the experimental energy relaxation rate significantly deviates downward from the theoretical value significantly. Such divergence for all the cases takes place when the conductivity of 2D gas falls below  $G_0$ (Fig.~\ref{f4}).

A possible reason for such divergence could be the change of a conductivity mechanism (it is commonly believed that at $\sigma<\pi G_0$ conductivity is hopping), and in this case the described above treatment with $P$ is not valid any more. To clarify the conductivity mechanism let us analyze the derivative  $\frac{\partial P}{\partial T_h}$ \cite{PhysE}.


From Eq.~(\ref{eq1}) it is seen that in a diffusive regime the derivative $\partial P/\partial T_h$ is independent of lattice temperature $T_L$. It means that experimental dependences $\frac{\partial P}{\partial T_h}=f(T_h)$ , obtained at different $T_L  $ should fall on the universal curve\cite{PhysE}. This statement is valid when the change of $\sigma$ with the electric field originates from the change of the hole temperature only. As the consequence, the set of curves $\partial P/\partial \sigma$ as a function of $\sigma$ also has such a property. Besides, the treatment with $\partial P/\partial \sigma$ is more consequent because when the approximation of the hole temperature fails, $\partial P/\partial \sigma$ remains defined.

In the hopping regime, firstly, $\sigma$ depends both on lattice and hole temperatures. Secondly, the change of conductivity with the electric field results not only from hole heating, but also from the change in probability of hopes. Finally, the energy distribution function of holes in the electric field can deviate from the Fermi-Dirac function. All these effects have to lead to a divergence of dependences $\frac{\partial P}{\partial \sigma}=f(\sigma)$ at different $T_L$.

\begin{figure}
\includegraphics[width=0.8\linewidth,clip=true]{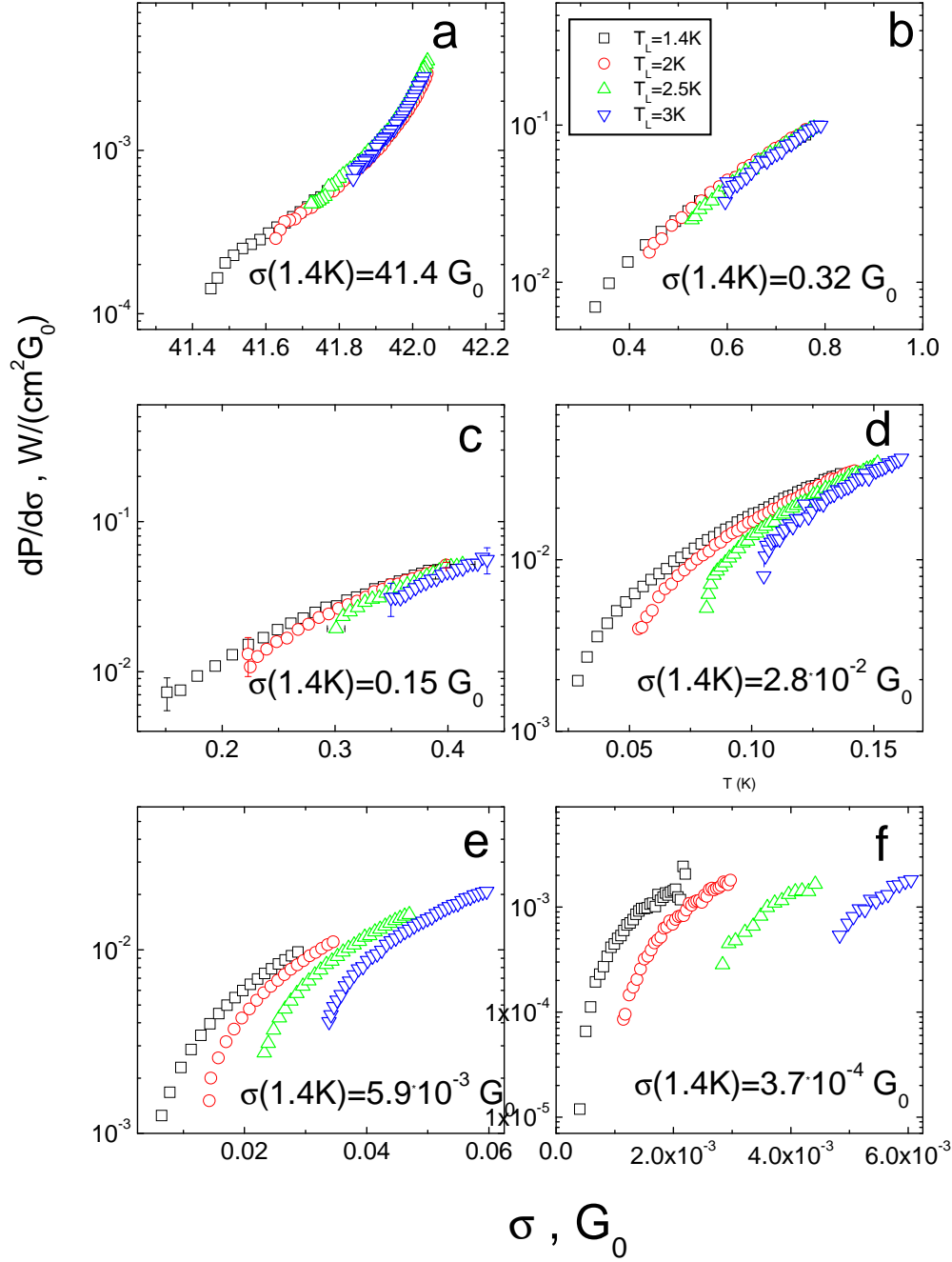}
\caption{(Color online) The derivative of energy relaxation rate with respect to $\sigma$  at different lattice temperatures for sample 3855. The values of conductivities at $T_L=1.4K$ are presented at figures.
}\label{f5}
\end{figure}

Dependencies $\partial P/\partial \sigma$ for sample 3855, obtained experimentally, are depicted in Fig.~\ref{f5}. As seen from Fig.~\ref{f3}(a) the data taken at different lattice temperatures fall on one universal dependence when the conductivity of the 2D gas is relatively high $\sigma(1.4K)=44.1G_0$ . With decreasing of conductivity this behavior remains till the conductivity reaches the value of  $2.8\cdot10^{-2}G_0$ (Fig.~\ref{f3} (b,c)). And only at $\sigma \approx 2.8\cdot 10^{-2}G_0\approx 10^{-2}\frac{e^2}{h}$ the curves begin to diverge, and with a decrease in conductivity they diverge drastically (Fig.~\ref{f3} (d,e,f)). Such divergence indicates a crossover to a hopping regime of conductivity in the investigated structures.

\begin{figure}
\includegraphics[width=0.8\linewidth,clip=true]{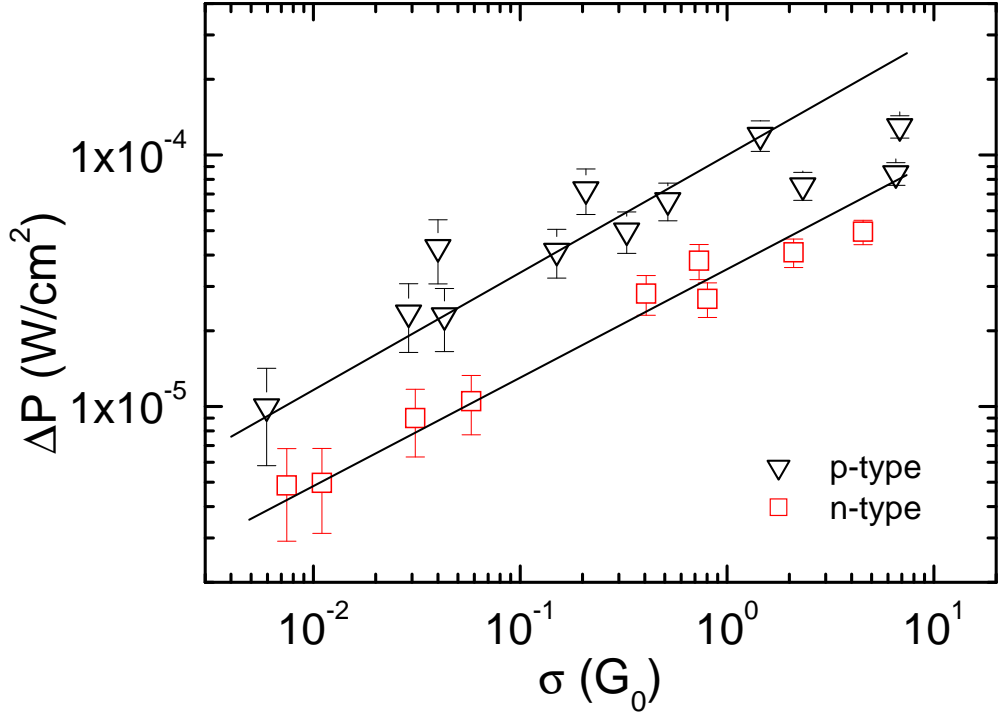}
\caption{(Color online) The dependence of power $\Delta P$ required to heat holes from lattice temperature $T_L=1.4K$ up to $T_h=1.9K$ as a function of conductivity for p-type and n-type structures.
}\label{f6}
\end{figure}

Hence, the conductivity remains diffusive at $\sigma(1.4K)>3\cdot 10^{-2}G_0$ and the drop in the energy relaxation rate at conductivity range $3\cdot 10^{-2}G_0<\sigma<1G_0$ is caused not by a crossover to the hopping conductivity. We believe that the it is caused by the Pippard ineffectiveness of the electron-phonon interactions\cite{Pippard}, which takes place under the following conditions: i) the number of carriers within the length of the thermal phonon is sufficient to introduce the local conductivity $\frac{q_t}{k_F}<1$ ($q_t$ is the wave vector of the thermal phonon), and ii) $q_tl<1$, ($l$ means free path). It was shown in Ref.~\onlinecite{Reizer}, that the energy relaxation rate would decrease linearly in this regime. Indeed (see Figs.~\ref{f4},\ref{f6}), in the investigated structures the linear decrease of the energy relaxation rate is observed. It begins at $\sigma(1.4K)\approx G_0$ , when both conditions mentioned above are satisfied: $q_tl\approx 0.2$ ($q_tl=1$ at $\sigma \approx 8G_0$ ). The parameter $\frac{q_t}{k_F}$ remains smaller than the unity the within whole range of temperatures. It should be noted that the analogous behavior was observed also on n-type structures (Figs.~\ref{f6})\cite{ntype}.

\section{Conclusion}
We have shown that in the diffusive regime at conductivities above $G_0$ the energy relaxation rate of 2D holes structures is well described in terms of inelastic scattering on acoustic phonons with both piezoelectric and deformational coupling to holes. It was shown that within the conductivity range of $(3\cdot 10^{-2}-1)G_0$  the conductivity remains diffusive, while the energy relaxation rate deviates from the theoretical prediction of \onlinecite{Price} downwards. Such a linear decrease results from the Pippard ineffectiveness for case $q_tl<1$, $\frac{q_t}{k_F}<1$. The analysis of $\partial P/\partial \sigma$ at a different lattice temperature shows that below $10^{-2}G_0$ the conductivity is hopping.

\section*{Acknowledgments} 
This work has been supported in part by the RFBR (Grants No.~08-02-00662, 08-02-91962, 09-02-789, 09-02-12206, 10-02-00481, 10-02-91336).
We would like to thank E. Romanova for useful discussions.


\begin{thebibliography}{21}
\bibitem{fnt1} The estimation shows that under our conditions the energy relaxation time within the hole system is significantly shorter than the electron-phonon interaction time, thus, the carrier distribution function remains Fermi function with effective hole temperature $T_h$. Therefore, the treatment, we apply to the data is valid.
\bibitem{ftn2} It should be mentioned, that the authors got a better agreement of experimental data with the theory \onlinecite{Price}, using numerical coefficients from Refs. \onlinecite{Chow96}-\onlinecite{Ma}, obtained for electrons, while the samples were specified in \onlinecite{Gao} \onlinecite{Gao2} as p-type.
\end{thebibliography}
\end{document}